\documentstyle[prl,multicol,amssymb,aps,epsf]{revtex}

\begin{document}

\author{T. A. S. Haddad{$^\ast$}, S. T. R. Pinho{$^\dagger$} 
and S. R. Salinas}
\address{Instituto de F\'{\i}sica, Universidade de S\~{a}o Paulo\\
Caixa Postal 66318, 05315-970, S\~{a}o Paulo, SP, Brazil}
\title{Universal critical behavior of aperiodic ferromagnetic models}
\date{\today}
\maketitle

\vspace*{-0.5truecm}

\begin{abstract}

We investigate the effects of geometric fluctuations, associated with
aperiodic exchange interactions, on the critical behavior of $q$-state
ferromagnetic Potts models on generalized diamond hierarchical lattices. For
layered exchange interactions according to some two-letter substitutional
sequences, and irrelevant geometric fluctuations, the exact recursion
relations in parameter space display a non-trivial diagonal fixed point that
governs the universal critical behavior. For relevant fluctuations, this fixed
point becomes fully unstable, and we show the apperance of a two-cycle which
is associated with a novel critical behavior. We use scaling arguments to
calculate the critical exponent $\alpha$ of the specific heat, which turns
out to be different from the value for the uniform case. We check the
scaling predictions by a direct numerical analysis of the singularity of the
thermodynamic free-energy. The agreement between scaling and direct
calculations is excellent for stronger singularities (large values of $q$).
The critical exponents do not depend on the strengths of the exchange
interactions.
\end{abstract}

\vspace*{-1.0truecm}

\begin{multicols}{2}

Quenched disorder may change the critical behavior of ferromagnetic spin
systems\cite{harris}. Although disorder provides the more usual examples,
there are alternative ways to break translational invariance. For instance,
the interest in the study of quasicrystals\cite{xextman} inspired a number
of proposals of deterministic schemes to build spatially aperiodic
structures. In analogy with the Harris criterion to account for the effects
of quenched disorder, Luck\cite{luck} developed a heuristic reasoning to
gauge the relevance of geometric fluctuations (associated with aperiodic
interactions) on the critical behavior of ferromagnetic models on Bravais
lattices. In a recent publication\cite{mst}, Luck's criterion has been
exactly derived for a $q$-state ferromagnetic Potts model on a diamond-type
hierarchical lattice. Now we revisit this problem to show the existence of
an attractor that gives rise to a novel class of critical behavior.

Many investigations of aperiodic classical\cite{achiam} and quantum\cite
{luck2} Ising chains, and two-dimensional Ising models with aperiodic
layered interactions\cite{layered}, use the formalism of substitution rules
on alphabets for the construction of infinite aperiodic sequences of
elements, which are then associated with coupling constants along a
direction of the lattice. We have taken advantage of the structure of
hierarchical lattices\cite{berker} to build layered aperiodic models\cite
{physica a} which are then amenable to simple (and exact)
renormalization-group calculations. In parameter space, there is always a
``diagonal'', non-trivial, fixed point of the renormalization-group
recursion relations. For relevant geometric fluctuations, this diagonal
fixed point, associated with the critical behavior of the uniform (equal
couplings) model, becomes fully unstable, and therefore cannot be reached
from any nonuniform initial conditions in parameter space\cite{mst}. The
critical behavior of the aperiodically perturbed system, in case of a phase
transition, should then be governed by another attractor in parameter space.

Relevant aperiodicity is believed to drastically weaken, or eventually 
suppress, the critical singularities\cite {luck2}. Recently, convincing
Monte Carlo evidence\cite{berches} has been
presented to indicate that relevant layered aperiodicity in the $8$-state
Potts model on the square lattice drives the phase transition from first to
second-order (with critical exponents independent of the strength of the
couplings). In the present work, we also detect a weakening of the critical
singularities for relevant geometric fluctuations. Besides being
numerically exact, our results can be regarded as equivalent to a 
Migdal-Kadanoff approximation for the analogous problems on Bravais lattices, 
and may thus be put in perspective with the Monte Carlo findings\cite{berches}.

\begin{figure}
\begin{center}
\epsfbox{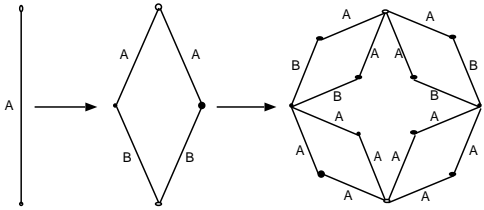}
\end{center}
\caption{Three successive generations of the 
simple diamond lattice ($b=m=2$).}  
\label{fig:fig1}
\end{figure}

The $q$-state Potts ferromagnet is defined by the Hamiltonian 
\begin{equation}
{\mathcal{H}}=-q\sum_{(i,j)}J_{i,j}\delta _{\sigma _{i},\sigma _{j}}, 
\label{hpotts}
\end{equation}
where $\sigma _{i}=1,2,\ldots,q$ for all sites of a lattice, $J_{i,j}>0$, and
the sum over $(i,j)$ refers to nearest-neighbor pairs of sites. We assume
that the couplings can take only two values, $J_{A}$ and $J_{B}$, associated
with the sequence of letters $A$ and $B$ of an aperiodic substitutional
word. To generate this sequence, we can use, for example, the successive
iterations of a period-doubling rule, given by $(A,B)\rightarrow (AB,AA)$.
In Fig.\ref{fig:fig1}, we show a simple diamond hierarchical lattice with 
aperiodic interactions according to this sequence (the letters, and the 
corresponding coupling constants, are chosen to mimic a layered structure). 
In general, we may consider basic ``diamonds'' with $m$ branches and $b$ bonds
along each branch, and generate hierarchical lattices with intrinsic, or 
fractal, dimension, $D=\ln (mb)/\ln (b)$. In each one of these structures,
aperiodicity may be implemented by a substitution rule of the form $
(A,B)\rightarrow (A^{n_{1}}B^{b-n_{1}},A^{n_{2}}B^{b-n_{2}})$, with $0\leq
n_{1},n_{2}<b$. These sequences are characterized by a substitution matrix
with eigenvalues $\lambda _{1}=b$ and $\lambda _{2}=n_{1}-n_{2}$. The
asymptotic form of the fluctuations in the number of letters depends on the
wandering exponent, 
\begin{equation}
\omega =\frac{\log \left| \lambda _{2}\right| }{\log \lambda _{1}}. 
\label{omega}
\end{equation}
Now we decimate the internal degrees of freedom of the diamonds to write
exact recursion relations for the reduced couplings. There is
always a non-trivial fixed point on the diagonal of parameter space, which
is associated with the critical behavior of the uniform model. This fixed
point becomes fully unstable for \cite{mst,physica a} 
\begin{equation}
\omega >1-\frac{D}{2-\alpha _{u}},  \label{criterio}
\end{equation}
where the critical exponent $\alpha _{u}$, associated with the specific heat
of the underlying uniform model, depends on $q$, $m$ and $b$ (note that the
transition is always continuous for the ferromagnetic Potts model on these
hierarchical lattices\cite{bambihu}). Therefore, Eq. (\ref{criterio}) is an
(exact) alternative statement of Luck's original heuristic criterion of
relevance of the geometric fluctuations. Irrelevancy of the fluctuations
corresponds to accessibility of the non-trivial symmetric fixed-point,
whatever the initial conditions. In the irrelevant cases, the critical
behavior of the aperiodic systems remains unchanged with respect to the
corresponding uniform models.

The problem now consists in the characterization of the new critical
behavior under relevant geometrical fluctuations. Andrade\cite{andrade} has
recently used a mapping for the succession of partition functions of
aperiodic Ising models ($q=2$) on hierarchical lattices to investigate this
question. For irrelevant geometric fluctuations, the critical behavior seems
to remain unchanged with respect to the uniform system. Although the results
are not conclusive, the critical behavior does seem to change for an Ising
model on a diamond-type lattice with $m=b=3$, and layered aperiodic exchange
interactions according to the rule $(A,B)\rightarrow (ABB,AAA)$. To go
beyond this work, we analyzed some families of aperiodic $q$-state Potts
ferromagnets on diamond-like lattices to show that they do present a novel
critical behavior (different from the uniform case, but universal) under
relevant geometric fluctuations. This novel critical behavior can be exactly
traced to new features in parameter space, exclusively related to the
relevant geometric fluctuations.

Consider the Potts model on a lattice with $b=2$ bonds per branch, and
interactions according to the period-doubling rule. From Eq.(\ref{criterio}), 
for $m=2$, fluctuations are relevant if $q>4+2\sqrt{2}$ (for $m=3$, the
threshold value of $q$ turns out to be somewhat lower, and so on). The
recursion relations are given by
\begin{mathletters}
\label{recdup}
\begin{equation}
x_{A}^{\prime }=\left( \frac{x_{A}x_{B}+q-1}{x_{A}+x_{B}+q-2}\right)^{m},
\end{equation}
and
\begin{equation}
x_{B}^{\prime }=\left( \frac{x_{A}^{2}+q-1}{2x_{A}+q-2}\right) ^{m},
\end{equation}
\end{mathletters}
where $x_{A,B}=\exp (q\beta J_{A,B})$, with $\beta =1/k_{B}T$. Besides the
uniform (diagonal) fixed-point, we have found a cycle of period two in the
relevant regime (that is, the two-cycle is present when the non-trivial
diagonal fixed point is fully unstable). Moreover, the two-cycle appears
clearly as a bifurcation, its points running away from the symmetric fixed
point as $q$ increases. In terms of the second iterate of
the recursion relations (of which each point of the two-cycle is a fixed
point), it displays a saddle-point character, with stable and unstable
manifolds. Supposing that this two-cycle is associated with a novel
critical behavior, we use standard scaling arguments to predict the critical
exponent $\alpha $ of the specific heat. Let us call $x$ the single relevant
variable describing the parameter space in the
neighborhood of the two-cycle. In the thermodynamic limit, we write the
reduced free energy per bond in the scaling form\cite{derrida},
\begin{equation}
f(x)=g(x)+\frac{1}{b^{2D}}f(x^{\prime \prime}),  \label{scalingform}
\end{equation}
where $g(x)$ is a regular function related to the free energy of the
decimated spins, $x^{\prime \prime }$ is a second iterate of the recursion
relations, and $b$ is the linear rescaling factor of the
renormalization-group transformation (which coincides, for diamond-like
lattices, with the parameter $b$ of the basic diamonds\cite{melrose}). Note
the presence of the factor $b^{2D}$, related to the need of two iterations
to go back to the neighborhood of the starting point in parameter space. 
Eq.(\ref{scalingform}) has the asymptotic solution
\begin{equation}
f(x)\simeq \left| x-x^{\ast }\right| ^{2-\alpha }P\left( \frac{\ln \left|
x-x^{\ast }\right| }{\ln \Lambda _{cic}}\right) ,  \label{formageraldef}
\end{equation}
where $x^{\ast }$ is one of the points of the two-cycle, $\Lambda _{cic}$ is
the largest eigenvalue of the linearization of the second iterate of the
recursion relations about any one of the points of the cycle, $P(z)$ is an
arbitrary function of period one, and the critical exponent $\alpha $,
associated with the specific heat, is given by 
\begin{equation}
{\alpha}=2-2\frac{\ln b^{D}}{\ln \Lambda _{cic}}=2-2\frac{\ln (mb)}
{\ln \Lambda_{cic}}.  \label{alpha}
\end{equation}

The values of $\alpha$ predicted by Eq.(\ref{alpha}) are unequivocally
different from the values $\alpha _{u}$ for the specific heat exponent in
the uniform cases, as can readily be seen in Table \ref{table:table1}, for 
the simple diamond lattice, $m=b=2$, with layered exchange interactions 
according to the period-doubling sequence. As in the Monte Carlo simulations 
for the aperiodic $8$-state Potts model on the square lattice\cite{berches}, 
there is a clear weakening of the critical singularities due to the geometric 
fluctuations (a feature also present in disordered systems). 
Similar conclusions can be drawn for $b=2$ diamond-type 
lattices with different values of $m$ and $q$, and the period-doubling 
substitution.
\end{multicols}

\widetext
\begin{table}
\caption{Results for the location of the two-cycle, eigenvalues of the second
iterate of the recursion relations about the points of the two-cycle and
the specific-heat critical exponent, $\alpha$, as predicted by 
Eq.(\ref{alpha}), for some values of $q$, in the case of the $m=b=2$ diamond 
with the period-doubling substitution. The value of the exponent in the 
uniform $(J_{A}=J_{B})$ case, $\alpha_{u}$, is also shown for comparison.}
\begin{tabular}{cccdddd}
 &\multicolumn{2}{c}{Location of the two-cycle}&\multicolumn{2}{c}
{Eigenvalues of 2nd iterate}& &\\
{$q$}&{$(x_{A},x_{B})_{1}$}&{$(x_{A},x_{B})_{2}$}&{$\Lambda_{1}$}
&{$\Lambda_{2}$}&{$\alpha$}&{$\alpha_{u}$}\\ \tableline
7&(5.285\ldots,7.642\ldots)&(6.697\ldots,4.750\ldots)&3.993\ldots&0.985\ldots
&-0.0022\ldots&0.010\ldots\\
25&(6.942\ldots,234.34\ldots)&(39.023\ldots,3.831\ldots)&4.243\ldots
&0.343\ldots&0.0817\ldots&0.404\ldots\\
100&(181.71\ldots,5.721\ldots)&(13.753\ldots,5151.84\ldots)&4.975\ldots
&0.074\ldots&0.2720\ldots&0.648\ldots\\
\end{tabular}
\label{table:table1}
\end{table}

\begin{multicols}{2}
We performed the same analysis for the Potts model on the $b=3$ generalized
diamond lattice, with interactions according to the rule $(A,B)\rightarrow
(ABB,AAA)$. Now, the recursion relations are given by
\ifpreprintsty
\else 
\end{multicols}\vspace*{-3.5ex}{\tiny \noindent 
\begin{tabular}[t]{c|}
\parbox{0.493\hsize}{~} \\ \hline
\end{tabular}
} \fi

\begin{mathletters}
\label{partea}
\begin{equation}
x_{A}^{\prime}=\left( \frac{x_{A}x_{B}^{2}+(q-1)x_{A}+2(q-1)x_{B}+q(q-3)+2%
}{x_{B}^{2}+2x_{A}x_{B}+(q-2)x_{A}+2(q-2)x_{B}+q(q-3)+3}\right) ^{m},
\end{equation}
and 
\begin{equation}
x_{B}^{\prime }=\left( \frac{x_{A}^{3}+3(q-1)x_{A}+q(q-3)+2}{%
3x_{A}^{2}+3(q-2)x_{A}+q(q-3)+3}\right) ^{m}.  \label{parteb}
\end{equation}
\end{mathletters}
\ifpreprintsty 
\else
{\tiny\hspace*{\fill}\begin{tabular}[t]{|c}\hline
\parbox{0.49\hsize}{~} \\
\end{tabular}}\vspace*{-2.5ex}%
\begin{multicols}{2}\noindent
\fi

For all values of $m$, the condition of relevance, given by Eq.(\ref
{criterio}), is satisfied for $q\geq 2$ (including the Ising model, $q=2$).
Again, besides the fully unstable fixed point, we have detected the presence
of a two-cycle. The same sort of scaling analysis has been performed. 
Table \ref{table:table2} gives some results for $b=3$ and $m=2$. We see 
that the weakening of the critical singularities is again indicated by 
these data, the same trend being present for $m=3$.
\end{multicols}

\widetext
\begin{table}
\caption{Same as Table \ref{table:table1}, for the lattice with $b=3$ and 
$m=2$, with the rule $(A,B)\rightarrow (ABB,AAA)$.}
\begin{tabular}{cccdddd}
 &\multicolumn{2}{c}{Location of the two-cycle}&\multicolumn{2}{c}
{Eigenvalues of 2nd iterate}& &\\
{$q$}&{$(x_{A},x_{B})_{1}$}&{$(x_{A},x_{B})_{2}$}&{$\Lambda_{1}$}
&{$\Lambda_{2}$}&{$\alpha$}&{$\alpha_{u}$}\\ \tableline
2&(6.446\ldots,135.10\ldots)&(34.794\ldots,5.224\ldots)&3.255\ldots
&0.311\ldots&-1.0358\ldots&-0.902\ldots\\
7&(11.469\ldots,1649.66\ldots)&(126.59\ldots,8.525\ldots)&4.755\ldots
&0.075\ldots&-0.2981\ldots&-0.185\ldots\\
100&(57.223\ldots,1121275.02\ldots)&(3272.70\ldots,34.683\ldots)&9.808\ldots
&0.001\ldots&0.4304\ldots&0.898\ldots\\
\end{tabular}
\label{table:table2}
\end{table}

\begin{multicols}{2}
The presence of these two-cycles seems to be associated with the alternance
of two energy scales, given by $J_{A}$ and $J_{B}$. The iteration of the
recursion relations leads to the alternative dominance of each one of them
(as it can already be seen in the behavior of the fluctuations in the number
of letters $A$ and $B$ along a substitution sequence). To check this
argument, we performed some calculations for the ferromagnetic Potts model
on a simple diamond lattice, $m=b=2$, with aperiodic interactions according
to the four-letter Rudin-Shapiro rule, $(A,B,C,D)\rightarrow (AC,DC,AB,DB)$,
whose fluctuations are known to be relevant even in the Ising case\cite
{physica a}. Indeed, we have found a cycle of period four, along with fully
unstable two-cycles, in regions of parameter space associated with the
symmetries of the sequence.

To test the validity of the scaling arguments, and of the role of the
two-cycle as the responsible for the new critical behavior, we have
performed a direct numerical analysis of the singularity of the free energy.
In real-space renormalization-group calculations, it is well known that the
(reduced) free energy can be expressed as an infinite series\cite{nauenberg}. 
For the Potts model on the $b=2$ diamond-type lattice, with the
period-doubling rule, it takes the form 
\begin{eqnarray}
f(x_{A},x_{B})=\sum_{n=0}^{\infty}\frac{1}{(2m)^{n}}&&\left[\frac{1}{3}\ln 
\left( x_{A}^{(n)}+x_{B}^{(n)}+q-2\right)\right.\nonumber \\
& &+\left.\frac{1}{6}\ln \left(2x_{A}^{(n)}+q-2\right)\right],  \label{serie}
\end{eqnarray}
where $x_{A,B}^{(n)}$ indicate the $n$-th iterates of the recursion
relations, Eq.(\ref{recdup}). If we assume uniform convergence, this series
can be differentiated term-by-term, and then summed in a computer, to obtain
the specific heat per bond. We also used direct numerical differentiation as
a control of this assumption of uniform convergence. The critical
temperature can be determined with high precision by making use of the
existence of the trivial paramagnetic fixed point at $T=0$, corresponding to 
$x_{A,B}=\infty $, which causes the apparent divergence of the series (\ref
{serie}) if summed without the use of any regularization trick. Fixing the
parameters $q$ and $m$, and also the strengths of $J_{A}$ and $J_{B}$, the
critical temperature thus calculated in fact places $x_{A,B}$ in the
attraction basin of the two-cycle. For irrelevant aperiodicity, as well as
for the uniform model, this method yields a critical temperature that
locates the system on the stable manifold of the uniform fixed point.

The singularity in the specific heat can be analyzed by a (non-linear)
fitting of a function to the data over a somewhat arbitrary scaling region.
For the uniform and irrelevant cases, we have always obtained very good
fittings, in excellent agreement with the values of $\alpha _{u}$ predicted
by the usual scaling theory around the uniform fixed point. As it should be
anticipated, these fitted values did not present any detectable sensitivity
on the values of $J_{A}$ and $J_{B}$. The particular problem of the Ising
model ($q=2$) on the simple $m=b=2$ diamond lattice, with exchange
interactions according to a period-doubling sequence, had already been
studied by Andrade\cite{andrade}, with the same conclusions. In the relevant
cases, the situation is much subtler, and demands a more refined analysis.
For large values of $q$, in which cases the scaling analysis predicts
positive values of $\alpha$ (although, of course, smaller than the
corresponding values of $\alpha_{u}$), the fittings presented excellent
agreement with the scaling predictions. For weaker singularities (mainly 
$\alpha$ negative), the fitted values were always somewhat bigger than the
scaling predictions, with better agreement for increasing values of $q$. For 
$m=b=2$, and the period-doubling sequence, let us give some results of the
fittings of the specific heat data to a function of the form $A+B\left|
t\right|^{-\alpha}$, where $t$ is the reduced temperature, and the
parameters $A$ and $B$ must not be universal. For $q=7$, 
we obtained $\alpha=-0.005(4)$, which should be compared with $\alpha
=-0.0022\ldots$. For $q=25$, we have $\alpha=0.08(2)$, to be compared with the
scaling value $\alpha=0.0817\ldots$. For $q=100$, we have $\alpha=0.27(1)$,
to be compared with $\alpha=0.2720\ldots$. Even in the cases of disagreement
with the scaling predictions, the fittings indicate no sensitivity on the
particular values of $J_{A}$ and $J_{B}$, and thus characterize the
universal nature of the critical behavior. For weak singularities, the
discrepancies in the results cannot probably be explained in terms of
lack of numerical precision, although it is in fact difficult to obtain a
fully reliable fitting in these cases. This behavior can be probably traced
to corrections to scaling that we are not considering in the direct
application of the scaling ideas to the two-cycles. It should be pointed out
that the free energy takes different values in each point of the two-cycle,
in such a way that it may be ill-defined in terms of just a simple scaling
field.

We have performed the same kind of numerical check for a lattice with $b=3$,
and with the aperiodic rule $(A,B)\rightarrow (ABB,AAA)$. Now, aperiodicity
is already relevant for $q=2$ (Ising model), and the two-cycle can be found
for $q\geq 2$, and all values of $m$. The situation turns out to be exactly 
the same as before. The numerical value of the critical temperature indeed
locates the system on the attraction basin of the two-cycle. The agreement
between the numerical fittings and the scaling predictions for $\alpha $
improves as the critical singularity grows stronger, which happens with
increasing values of $q$. In the special case $q=2$, and $m=b=3$, we and
Andrade\cite{andrade} have calculated similar values, $\alpha =-0.90(9)$, to
be compared with the scaling prediction, $\alpha =-0.9684\ldots$.

From the numerical calculations, we have observed an oscillatory behavior 
in the specific heat as a function of temperature above $T_{c}$. The period of
these oscillations is roughly given by Eq.(\ref{formageraldef}), with better
agreement for increasing values of $q$. These oscillations have also been
found by Andrade\cite{andrade}, and are well known phenomena related to
hierarchical structures\cite{oscil}.

In conclusion, we have given a number of examples of ferromagnetic Potts
models on diamond-type hierarchical lattices to show that irrelevant
geometric fluctuations do not change the (universal) critical behavior with
respect to the uniform cases. On the other hand, for two-letter
substitutional sequences, relevant geometric fluctuations give rise to a
novel universal critical behavior associated with a two-cycle in parameter
space.

We thank discussions with R. F. S. Andrade and very helpful suggestions and
comments by A. P. Vieira and R. M. Dami\~{a}o. This work has been supported
by the Brazilian agencies FAPESP, CAPES, and CNPq.

\end{multicols}

\end{document}